\documentclass[journal]{IEEEtran}

\usepackage{myStyleIEEE}
\usepackage{nomencl}
\usepackage{ragged2e}
\usepackage[thinc]{esdiff}
\IEEEoverridecommandlockouts

\newlength{\bracewidth}

\usepackage{etoolbox}
\renewcommand\nomgroup[1]{%
  \item[\bfseries
  \ifstrequal{#1}{P}{A. Parameters}{%
  \ifstrequal{#1}{V}{C. Variables}{%
  \ifstrequal{#1}{S}{B. Sets and Indices}{}}}%
]}

\begin{document}

\title{Stability Constrained Optimization in High IBR-Penetrated Power Systems{\textemdash}Part I: Constraint Development and Unification}

\newtheorem{proposition}{Proposition}
\renewcommand{\theenumi}{\alph{enumi}}

\author{Zhongda~Chu,~\IEEEmembership{Member,~IEEE,} and
        Fei~Teng,~\IEEEmembership{Senior Member,~IEEE} 
        
        
\vspace{-0.5cm}}
\maketitle
\IEEEpeerreviewmaketitle

\begin{abstract}
Conventional power system optimization framework is becoming less reliable and efficient due to the stability issues brought by the ever-increasing inverter-interfaced renewable penetration. To ensure system stability during system operation and to provide appropriate incentives in the future market-based stability maintenance framework, it is essential to develop a comprehensive set of power system stability constraints which can be incorporated into system optimization. In this paper, different system stability issues, including synchronization, voltage and frequency stability, are investigated and the corresponding stability conditions are analytically formulated as system operational constraints. A unified framework is further proposed to represent the stability constraints in a general form and enable effective reformulation of the impedance-based stability metrics. All the constraints are converted into linear or Second-Order-Cone (SOC) form, which can be readily implemented in any optimization-based applications, such as system scheduling, planning and market design, thus providing significant value for multiple system stability enhancement and studies.
\end{abstract}


\makenomenclature
\renewcommand{\nomname}{List of Symbols}
\mbox{}
\nomenclature[P]{$M$}{system inertia$\,[\mathrm{MWs/Hz}]$}

\section{Introduction} \label{sec:1}
Renewable Energy Sources (RES) have been massively integrated into the modern electric power system in the past few decades due to the concerns on environment and sustainability throughout the world. Being the key element for the interface between RES and the grid, the power electronic converters are anticipated to acquire a steadily increasing role along the trend of decarbonization. However, owing to the intermittent nature of renewable energy and the distinguished features of Inverter-Based Resources (IBR), which stem from their control interactions, zero inertia provision, limited Short Circuit Current (SCC) and reactive power supports, power systems are facing new challenges in system operation, security and stability \cite{8450880,Blaabjerg2006,Ulbig2014,Carrasco2006}. 

Among all these challenges, the frequency and inertia issues were first reported and have drawn significant attention from the researchers\cite{8450880,TIELENS2016999}. \textcolor{black}{RES are connected to the grid through power electronics inverters. Due to their asynchronous nature, replacing the conventional SGs with IBR would gradually lead to low-inertia power systems, whose frequency stability may be jeopardized. Another issue due to the high IBR penetration is the voltage stability and reactive power (current) supply in both transient and static stages. This is because of the limited SCC contribution from IBR during transient processes, around 1-2 p.u. \cite{7878663} compared with conventional SGs (5-10 p.u.\cite{TLEIS2019597}) and the current source control without direct voltage support during normal operation. Challenges in the synchronization stability of IBR have also been identified in high IBR-penetrated systems where the system strength at IBR buses is low, due to the existence of Phase-Locked Loops (PLLs)\cite{9181463,6848832}.}

Measures in different respects have been taken to address the issues associated with different stability issues, such as \cite{8579100,9449821} for frequency stability, \cite{9716751,8728057} for voltage stability and \cite{9201068,9082101} for synchronization stability. Most of the existing research tends to resolve those issues from the perspective of device-level control design, where the processes of parameter tuning and the assessment of the stability of the entire system can be difficult. Moreover, solely depending on device-level control may fail to maintain system stability, especially during the transient processes where the IBR currents reach their limits, and the optimal economic performance may not be achieved without system-level regulations. For example, to ensure frequency security in the most cost-effective manner, different frequency services need to be coordinated and co-optimized during the system scheduling process, while considering the device-level control strategies. Therefore, optimal solutions need system-level coordination of different resources to ensure various stability and maximize economic profit. 

To achieve this purpose, it is necessary to develop system stability constraints, which can be further incorporated into system-level optimizations. Although some work has been done on the system level to address stability issues such as \cite{9925092,7399774,9066910} on frequency and \cite{2020JEET,app9163412,9786660} on voltage stability, only a specific stability issue is studied and the interactions among different stability are not investigated due to the complicated dynamics of different stability phenomena. There still lacks a unified stability-constrained optimization framework in power systems which can simultaneously incorporate various stability constraints in high IBR-penetrated systems. 

Another challenge preventing stability constraints from being embedded into power system optimization models is the reformulation of complex stability criteria. Typically, the criteria for different stability are highly nonlinear or even derived without explicit expressions, and hence cannot be directly incorporated as constraints into system optimization models. To solve this problem, analytical approaches such as piecewise linearization have been applied to transform the nonlinear stability criteria into linear form \cite{8667397,9066910}. However, due to the intricate system dynamics, these methods may lead to over-conservative results, compromising the economic benefit of the system. 

In this context, this paper focuses on power system stability issues due to high IBR penetration, including frequency stability, synchronization stability, and voltage stability. Stability constraints that depend on the system steady-state operating conditions are developed and a unified framework is proposed to reformulate the highly nonlinear constraints into linear or SOC form by combining analytical and data-driven approaches. Note that the newly extended converter-driven and resonance stabilities \cite{9286772} are not covered in this work, as on the one hand, with limited research in this area, the system-level indicators of these two types of stability are not clear and on the other hand, they are more relevant to converter control design and less to the system level conditions. The main contributions of this paper include the following:
\begin{enumerate}
    \item Different stability criteria in high IBR-penetrated power systems are analytically formulated as operational constraints within the same framework, allowing the coordination of different resources for various stability maintenance and the investigation of the interaction between different stability constraints. Depending on system impedance, power injections, and generator status, these constraints cover frequency stability, synchronization stability and voltage stability and can be applied to any power system optimization model, such as system scheduling, planning and market design. Contributions are also made during the derivation of the following two stability constraints.
    \begin{itemize}
        \item Synchronization stability of PLL-interfaced grid-following (GFL) IBRs is formulated as operational constraints and incorporated into the proposed framework, to ensure both the existence of equilibrium point and the small-signal synchronization stability.
    
        \item Transient voltage stability constraints are developed based on short circuit currents and post-fault voltages. An SCC quantification method is proposed where the IBRs during grid fault are modeled as voltage-dependent current sources to account for the actual voltage drop at IBR terminals. 
    
    \end{itemize}

    \item A novel unified reformulation framework is proposed, which transforms the derived power system stability constraints into SOC form. It combines analytical derivation and a data-driven approach to achieve both conservativeness and accuracy. 
\end{enumerate}
    
It should be noted that the proposed stability-constrained framework does not intend to replace the stability assessment based on more accurate system modeling or time-domain simulations, which are currently carried out in near-real time after system-level optimization to ensure system stability. Instead, with the stability constraints in the optimization model, the number of generation adjustments after the near-real time stability assessment can be significantly reduced compared with stability-unconstrained optimization, thus increasing the efficiency and economic benefit of system operation. 

The rest of the paper is organized as follows. In Section~\ref{sec:2}, different system stability constraints are developed, including synchronization and voltage stability of GFL IBRs, and system frequency stability. The resulting stability constraints are further converted to SOC or linear forms in Section~\ref{sec:3}  which can be directly included in power system optimization model. Section~\ref{sec:4} concludes the paper.

\section{Power System Stability Constraint Development} \label{sec:2}
In this section, different stability constraints are derived analytically. Specifically, we focus on the synchronization and voltage stability of GFL IBRs, and system frequency stability. For the synchronization stability, the existence of equilibrium point, small-signal and transient stability are discussed, whereas for the voltage stability, both small-signal and transient stability are investigated. The frequency stability is assessed based on the security constraints specified by system operators, such as maximum Rate of Change of Frequency (RoCoF), frequency nadir and steady-state frequency. It should be noted that the focus of this work does not lie in proposing new stability indices but in developing system-level operational constraints and further converting them into optimization-friendly forms through a unified reformulation framework. Although specific stability indices are chosen in this section, the proposed method can also be applied to others.

Consider a power system having $n\in\mathcal{N}$ buses with $g\in\mathcal{G}$, $c_l\in\mathcal{C}_l$ and $c_m\in\mathcal{C}_m$ being the set of conventional Synchronous Generators (SGs), GFL and grid-forming (GFM) inverter-based generators. $\Psi(g)$ and $\Phi(c)$ map the units in $g\in \mathcal{G}$ and $c\in \mathcal{C}=\mathcal{C}_l\cup\mathcal{C}_m$ to the corresponding bus indices respectively. \textcolor{black}{For the stability analysis, the SGs are modeled as voltage sources behind a reactance whereas the GFL IBRs are modeled as controlled current sources. The GFM IBRs are modeled as voltage sources in small-signal analysis and as voltage-dependent current sources during transient due to the current saturation. Loads are modeled as constant current loads for synchronization and voltage stability \cite{yuan2022assessing,8488538}. More details regarding the IBR model can be found in Part~II of this paper.}


\subsection{Synchronization Stability of IBRs} \label{sec:2.1}
The IBR units are generally synchronized with the grid through one of the two methods, namely voltage-based grid synchronization (GFL control) and power-based synchronization (GFM control) \cite{9181463}. The former relies on the estimation or measurement of the frequency and phase of the voltage at the point of common coupling which is achieved through the implementation of PLL, whereas the latter directly controls the phase of PCC voltage by regulating the IBR active power, such as $P-f$ droop control. Different research suggests that the GFL IBRs may suffer from synchronization instability under weak connection with the grid, under both small and large disturbances. To ensure the synchronization stability of GFL IBRs, the constraints corresponding to the existence of equilibrium point, small-signal and transient stability are developed in this subsection.

\subsubsection{Existence of equilibrium point}
To ensure the synchronization stability of IBRs, there should always exist an equilibrium point for GFL IBRs. 

\begin{figure}[!t]
        \centering
    	\scalebox{1.5}{\includegraphics[trim=0 0 0 0,clip]{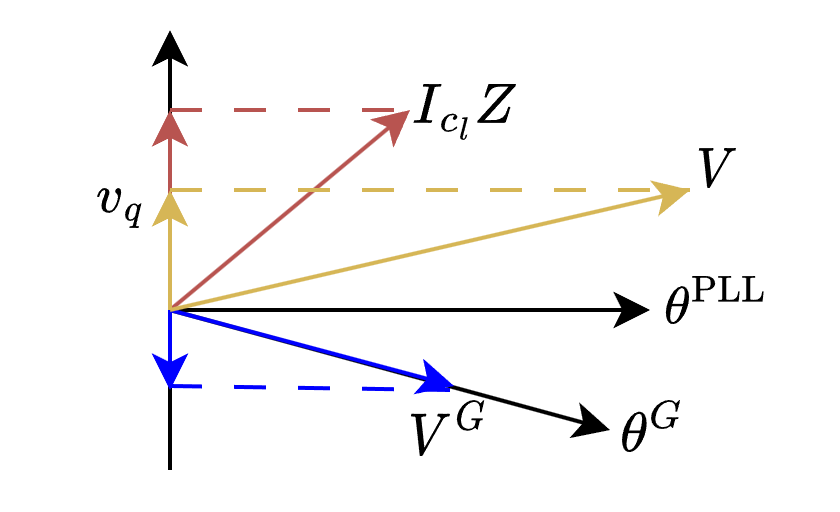}}
        \caption{\label{fig:Phasor}\textcolor{black}{Phasor diagram of PLL dyanmics}.}
        \vspace{-0.4cm}
    \end{figure}

The dynamics of a typical PLL structure in GFL IBRs can be described as:
\begin{subequations}
\label{eq:pll}
\begin{align}
    &\diff{\left(\theta^{\mathrm{PLL}}_{\Phi(c_l)}-\theta^G_{\Phi(c_l)}\right)}{t} = K_p v_q + \int_0^t K_i v_q \, \mathrm{d}t \label{pll_1} \\
    &v_q = -|V^G_{\Phi(c_l)}|\sin{\left(\theta^{\mathrm{PLL}}_{\Phi(c_l)}-\theta^G_{\Phi(c_l)}\right)} \nonumber\\
    & \quad + |I_{c_l}Z_{\Phi(c_l)\Phi(c_l)}|\sin{\left(\phi_{c_l}+\phi_Z^G\right)}, \label{pll_2}
\end{align}
\end{subequations}
where $\theta^{\mathrm{PLL}}_{\Phi(c_l)}$ and $\theta^G_{\Phi(c_l)}$ are the phase angles of PLL and grid voltage $V^G_{\Phi(c_l)}$; $v_q$ is the q-axis component of PCC voltage $V_{\Phi(c_l)}$; $K_p$ and $K_i$ are the PI control gains; $\phi_{c_l}$ and $\phi_Z^G$ denote the angle of IBR output current $I_{c_l}$ defined on the PLL frame, and the angle of grid impedance $Z_{\Phi(c_l)\Phi(c_l)}$. \textcolor{black}{The relationship between quantities is shown in Fig.~\ref{fig:Phasor}, where the subscript $\Phi(c_l)$ is omitted for clarity. Note that \eqref{pll_2} can be derived from: $V_{\Phi(c_l)} = V^G_{\Phi(c_l)} + I_{c_l}Z_{\Phi(c_l)\Phi(c_l)}$}

At the equilibrium point of \eqref{eq:pll}, $\diff{\theta^{\mathrm{PLL}}}{t} \equiv f^{\mathrm{PLL}}=f^G \equiv \diff{\theta^{G}}{t}$, i.e., the PLL frequency equals the grid frequency and the GFL IBR is synchronized with the grid. Therefore, for stable operation, the PLL must be able to regulate $v_q$ in \eqref{eq:pll} to zero, leading to one necessary condition (the existence of equilibrium points) of IBR synchronization stability:
\begin{equation}
\label{eq:sync_gfl_I}
    I_{c_l} \le \frac{V^G_{\Phi(c_l)}}{Z_{\Phi(c_l)\Phi(c_l)} \sin{(\phi_{c_l}+\phi_Z^G)}}.
\end{equation}
In order to derive system-level operational constraints, the angle $\phi_{c_l}$ is eliminated with by the active and reactive power injections $P_{c_l},\,Q_{c_l}$: 
\begin{equation}
\label{eq:sync_gfl_PQ}
    \frac{V^G_{\Phi(c_l)} V_{\Phi(c_l)}}{Z_{\Phi(c_l)\Phi(c_l)}} \ge \cos \phi_Z^G Q_{c_l} + \sin{\phi_Z^G} P_{c_l}.
\end{equation}
The above constraint is a necessary condition for IBR synchronization stability and should apply during both normal operation and grid fault. \textcolor{black}{Since the focus of this work is the stability-constrained optimization during steady-state operation, which can be viewed as preventive control, the existence of equilibrium points during transient processes is not implemented as operational constraints. Instead, it can be maintained with the proper selection of the IBR control method after faults. More discussions are given in Section~\ref{sec:2.1.3}. Moreover, \eqref{eq:pll} also indicates that the control gains $K_p$ and $K_i$ influence the stability (both small-signal and transient). However, these control parameters are tuned and optimized during the IBR control design with the consideration of the overall IBR performance and stability. It may not be practical to change these parameters frequently as decision variables in the steady-state system-level optimization. Therefore, these control gains are treated as fixed parameters when developing the stability constraints.}

\subsubsection{Small signal synchronization stability}
In weak grids (large grid impedance), the IBR PCC voltage can be significantly affected by the current injection into the grid, forming a self-synchronization loop (positive feedback) and hence undermining the GFL IBR synchronization stability. The method proposed in \cite{8488538,gOSCR} is utilized here where the small signal stability of PLL-based GFL IBRs is assessed through the generalized short circuit ratio (gSCR). \textcolor{black}{The small-signal synchronization stability is dominated by the dynamics of IBRs and the network with which they are connected. Therefore, the closed-loop linearized system dynamics is first built by combining the dynamics of individual IBR and the network dynamics (dominated by the admittance/impedance matrix). Next, the small-signal stability analysis is carried out based on the linear control theory.} The definition of $Y_{eq}$ is given by:
\begin{subequations}
\begin{align}
    \mathrm{gSCR} &= \lambda_{\mathrm{min}} (Y_{eq}) \\
    Y_{eq} &= \mathrm{diag}\left(\frac{V^2_{\Phi(c_l)}}{P_{c_l}}\right) Y_{red} \label{Yeq},
\end{align}
\end{subequations}
where $\mathrm{diag}\left({V^2_{\Phi(c_l)}}/{P_{c_l}}\right)$ is the diagonal matrix of the GFL IBR terminal voltage and output power; $Y_{red}$ is the reduced node admittance matrix after eliminating passive buses and infinite buses. Note that $Y_{eq}=\mathrm{diag}\left(\frac{V^2_{\Phi(c_l)}}{P_{c_l}}\right) Y_{red}$ is diagonalizable with its smallest eigenvalue $\lambda_{\mathrm{min}} (Y_{eq})\in \mathbb{R}^+$ representing the connectivity of the network, and thus the grid voltage strength and the small-signal synchronization stability constraint can be then formulated as \cite{gOSCR}:
\begin{equation}
    \mathrm{gSCR} \ge \mathrm{gSCR}_{\mathrm{lim}},
\end{equation}
where $\mathrm{gSCR}_{\mathrm{lim}}$ is the critical (minimum) gSCR that needs to be maintained to ensure the small signal stability of the GFL units. Furthermore, based on the assumption that the system voltages stay close to $1\,\mathrm{p.u.}$ during normal operation and small disturbances, the critical gSCR is an operation-independent value, which can be determined offline with or without the detailed control parameters of the grid-following IBRs \cite{gOSCR}.

\textcolor{black}{The above small signal stability accounts for the impact of GFL (dynamics and locations) and the networks. The GFL dynamics influences the value of $\mathrm{gSCR}_{\mathrm{lim}}$, whereas the GFL capacity and location as well as the network influence the value of $\mathrm{gSCR}$, by influencing $Y_{eq}$. In other words, the gSCR-based stability analysis decouples the dynamics-related quantities with the steady-state quantities ($Y_{eq}$). The former influences $\mathrm{gSCR}_{\mathrm{lim}}$ and is determined by the GFL dynamics and the network dynamics, which are fixed at the system operation stage once the GFL control algorithm and parameters are selected. The latter influences $\mathrm{gSCR}$ and is determined by the GFL capacity and location as well as the admittance matrix. As a result, by forcing the $\mathrm{gSCR}\le \mathrm{gSCR}_{\mathrm{lim}}$ during system operation, the small signal stability can be maintained.}

\textbf{Impact of GFM units:}
Being modeled and controlled as voltage sources, the GFM converters improve the system strength, thereby enhancing the small-signal stability of power systems with large-scale PLL-based GFL converters. GFM converters can be modeled as a voltage source in series with an admittance with a large magnitude, which can be treated similarly to conventional SGs for small-signal analysis (without saturation). Hence, the small-signal stability constraints considering the impact of GFM converters can be represented by \cite{9282199}:
\begin{equation}
\label{ssss_forming}
    \mathrm{gSCR} = \lambda_{\mathrm{min}} \left[\mathcal{R}_{\mathcal{C}_m}(Y_{eq})\right]\ge \mathrm{gSCR}_{\mathrm{lim}},
\end{equation}
where $\mathcal{R}_{\mathcal{C}_m}(Y_{eq})$ is a function to remove the rows and the columns corresponding to all GFM units in $\mathcal{C}_m$.

\textcolor{black}{Although the derivation in \cite{8488538,gOSCR} does not include SGs explicitly, it includes an arbitrary number of infinite buses (ideal voltage sources), which increase the grid strength in the system. Since an SG or a GFM can be modeled as a voltage source behind an impedance for small signal analysis due to its voltage source behaviors, the impact of SGs and GFMs can be modeled by augmenting the system admittance matrix with the internal impedance of SGs and GFMs \cite{xin2022many}, as demonstrated in \eqref{Y}. By doing so, a general multi-machine system can be converted to the formulation in \cite{8488538,gOSCR}.}

\subsubsection{Large signal synchronization stability}\label{sec:2.1.3}
In addition to the small signal instability caused by weak grid connections, the IBR synchronization stability during grid faults should also be maintained. Various control methods have been proposed to enhance the transient synchronization stability of GFL IBRs during grid faults. Generally, this can be achieved through either active power adjustment or PLL parameter modification. With the proper selection of these control schemes, the transient synchronization stability can be maintained during grid faults. For instance, by aligning the output current vector with the negative grid impedance, the transient stability can always be guaranteed with mathematical proof \cite{7915732}, \textcolor{black}{which is briefly introduced here. This method modifies the IBR current injection during faults according to the grid impedance angle, i.e., $\phi_{c_l} = \tan^{-1} \left(\frac{I_{c_l}^d}{I_{c_l}^q} \right)= - \phi_Z^G$. Based on this control law, the equilibrium point during faults is guaranteed to exist since the following condition always holds.
    \begin{equation}
        |I_{c_l}| \le \frac{|V^G_{\Phi(c_l)}|}{|Z_{\Phi(c_l)\Phi(c_l)}| \sin{(\phi_{c_l}+\phi_Z^G)}}
    \end{equation}
Furthermore, to evaluate the stability of (1), Lyapunov method is used with the following Lyapunov function:
    \begin{equation}
        V(\lambda, \sigma) = \frac{1}{2} \lambda^2 + K_i |V^G_{\Phi(c_l)}| (1-\cos{\sigma})
    \end{equation}
where $\lambda = \int K_i v_q$ and $\sigma = \left(\theta^{\mathrm{PLL}}_{\Phi(c_l)}-\theta^G_{\Phi(c_l)}\right)$. The derivative of the Lyapunov function, along the trajectories of the system, can be derived as: $\dot V(\lambda, \sigma) = -K_i K_p |V^G_{\Phi(c_l)}| \sin^2\sigma$. With $K_i>0$ and $K_p>0$, we can obtain $V(\lambda, \sigma)\ge 0$ and $\dot V(\lambda, \sigma) \le 0$, indicating the system \eqref{eq:pll} being globally stable during the transient process.} Since the above stability analysis holds regardless of the normal operating conditions, the transient stability constraints during system operation are not considered in this work.

\subsection{Voltage Stability} \label{sec:2.2} 
\begin{figure}[!b]
    \centering
    \vspace{-0.4cm}
	\scalebox{1}{\includegraphics[trim=7 0 0 0,clip]{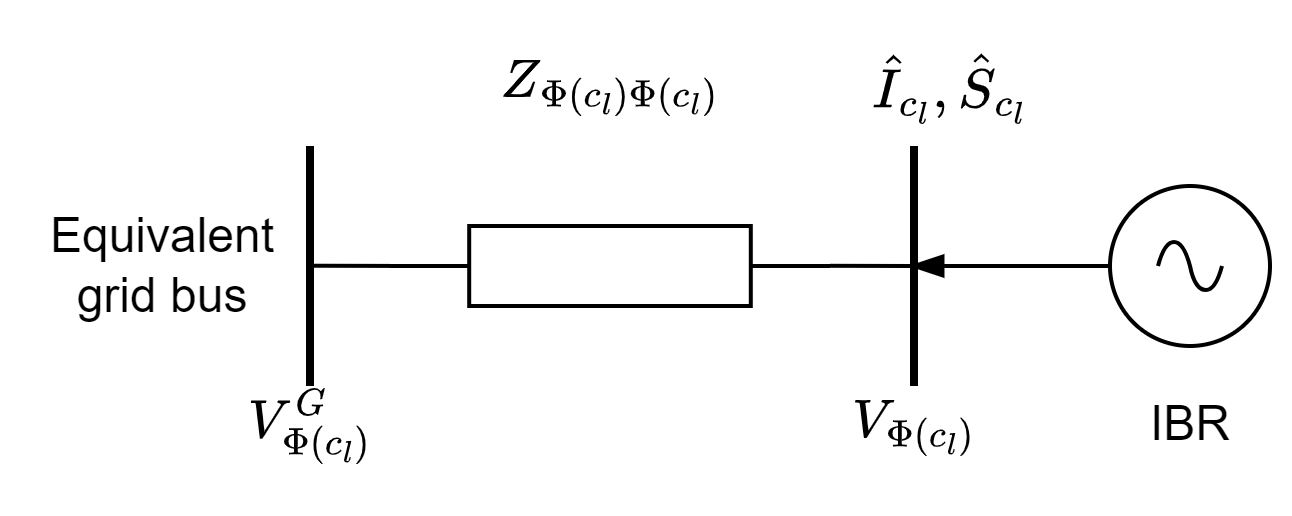}}
    \caption{\label{fig:2bus}Equivalent circuit of a general AC system seen from an IBR bus.}
\end{figure}

With the decline of conventional SGs and the corresponding reactive power and current support, system strength at the PCC of GFL IBRs decreases, which may bring voltage stability issues if left unmanaged. 
\subsubsection{Small signal (static) voltage stability}
The static voltage stability constraints previously derived in \cite{9786660} are briefly presented here. Based on the power flows in the system, the voltage and current relationship at the PCC of GFL IBR is expressed as:
\begin{align}
\label{V_c_eq}
    V_{\Phi(c_l)} & = \underbrace{\sum_{g\in\mathcal{G}\cup\mathcal{C}_m }Z_{\Phi(c_l)\Psi(g)} I_{g}}_{V^G_{\Phi(c_l)}} \nonumber \\*
    & + Z_{\Phi(c_l)\Phi(c_l)} \underbrace{\left(I_{c_l} + \sum_{c_l'\in\mathcal{C}, c_l'\neq c_l}\frac{Z_{\Phi(c_l)\Phi(c_l')}}{Z_{\Phi(c_l)\Phi(c_l)}}I_{c_l'}\right)}_{\hat I_{c_l}}.
\end{align}
The above relationship can be represented by an equivalent two-bus system with one bus being Bus $\Phi(c_l)$ and the other bus representing the rest of the system, i.e., the equivalent grid bus with its voltage $V^G_{\Phi(c_l)}$, formed by all the voltage sources in the system, as shown in Fig.~\ref{fig:2bus}. \textcolor{black}{Note that GFMs are treated the same as the SGs since their terminal voltages can be well regulated during steady state and small disturbances when the power and current saturation do not occur. For these PV buses, it is common in voltage stability analysis to assume that they have constant voltage phasors \cite{kessel1986estimating,wang2016necessary}. Furthermore, the formulation in \eqref{V_c_eq} is an alternative of interpreting the original system instead of a model simplification. A similar formulation can also be found in \cite{wu2017assessing}.} The GFL is connected to the grid bus through the impedance $Z_{\Phi(c_l)\Phi(c_l)}$ with an equivalent current injection $\hat{I}_{c_l}$, to account for the IBR interactions. This equivalent current includes the current from IBR $c_l$ and the currents from other IBRs referred to Bus $\Phi(c_l)$ by the impedance ratio. The equivalent complex power injection from IBR $c_l$ is then calculated by combining \eqref{V_c_eq} and $\hat{S}_{c_l} =\hat{P}_{c_l}+\mathrm{j} \hat{Q}_{c_l}$, which leads to:
\begin{align}
    \begin{bmatrix}
    \hat{P}_{c_l}\\ \hat{Q}_{c_l}
    \end{bmatrix} = \begin{bmatrix}
    \frac{|V_{\Phi(c_l)}||V^G_{\Phi(c_l)}|\sin \theta_{c_l}}{|Z_{\Phi(c_l)\Phi(c_l)}|} \\  {\frac{|V_{\Phi(c_l)}|^2-|V_{\Phi(c_l)}||V^G_{\Phi(c_l)}|\cos \theta_{c_l}}{|Z_{\Phi(c_l)\Phi(c_l)}|}}
    \end{bmatrix},
\end{align}
with $\theta_{c_l}=\angle{V_{\Phi(c_l)}} - \angle{V^G_{\Phi(c_l)}}$ being the angle difference between the PCC voltage and the equivalent grid voltage. Eliminating the angle $\theta_{c_l}$ gives the closed form solution of $V_{\Phi(c_l)}$:
\begin{align}
    |V_{\Phi(c_l)}|^2  = \frac{|V_{\Phi(c_l)}^G|^2}{2}+ \hat{Q}_{c_l}|Z_{\Phi(c_l)\Phi(c_l)}| \pm \sqrt{\Delta},
\end{align}
where 
\begin{equation}
    \Delta = \frac{|V_{\Phi(c_l)}^G|^4}{4} - \hat{P}_{c_l}^2|Z_{\Phi(c_l)\Phi(c_l)}|^2+\hat{Q}_{c_l}|Z_{\Phi(c_l)\Phi(c_l)}||V^G_{\Phi(c_l)}|^2.
\end{equation}
In order to ensure the voltage stability of the GFL IBR $c_l$, real roots of $|V_{\Phi(c_l)}|$ should exist, i.e., ${\Delta}\ge 0$, thus yielding the following voltage stability constraints in terms of the IBR maximum power transfer capability:
\begin{equation}
\label{VS}
    \hat{P}_{c_l}^2+\hat{Q}_{c_l}^2\le { \Bigg( \hat{Q}_{c_l}+ \underbrace{ \frac{{|V^G_{\Phi(c_l)}}|^2}{2|Z_{\Phi(c_l)\Phi(c_l)}|}}_{\Gamma_{c_l}} \Bigg) ^2},
\end{equation}
where $\hat P_{c_l}$ and $\hat Q_{c_l}$ take the form:
\begin{subequations}
\label{PQ_1}
\begin{align}
    \hat P_{c_l} &= P_{c_l}+   \sum_{c_l'\in \mathcal{C},c_l'\neq c_l}\frac{|Z_{\Phi(c_l)\Phi(c_l')}|}{|Z_{\Phi(c_l)\Phi(c_l)}|}P_{c_l'} \\
    \hat Q_{c_l} &= Q_{c_l}+   \sum_{c_l'\in \mathcal{C},c_l'\neq c_l}\frac{|Z_{\Phi(c_l)\Phi(c_l')}|}{|Z_{\Phi(c_l)\Phi(c_l)}|}Q_{c_l'}.
\end{align}
\end{subequations}
This voltage stability condition illustrates that the equivalent active power injection of IBR $c_l$ is limited by an upper bound depending on short circuit capacity $\left({{|V^G_{\Phi(c_l)}}|^2}/{|Z_{\Phi(c_l)\Phi(c_l)}|}\right)$ at Bus $\Phi(c_l)$ if no reactive power is provided by IBR. At weak points of the system where $|Z_{\Phi(c_l)\Phi(c_l)}|$ is large, the active power injection from IBRs may have to be set smaller than the rated or even the available power to maintain voltage stability. Furthermore, it is also shown in \eqref{VS} that properly setting GFL IBRs' reactive power injection would allow more active power to be transferred. \textcolor{black}{An electrically nearer IBR's reactive power has more support on the voltage stability of the IBR $c_l$, since the weighted factor is closer to 1, whereas for the IBR with a large electric distance, the weighted factor is close to zero, hence less contribution to the voltage stability. An SG or a GFM supports the voltage stability of the nearby buses $i\in \mathcal{I}$ by reducing the impedance $Z_{ii}$ (increasing the system strength) at bus $i$, since the SG/GFM is modeled as a voltage source in series with an impedance with a small magnitude.}

\textcolor{black}{Typically, the critical condition of voltage stability can be derived based on different methods, such as the power-voltage (`nose') curve, the voltage to power sensitivity ($d|V|/dP$) and the singularity of the power flow Jacobian matrix. All of them are derived based on the analysis of the steady-state power flow equations and are equivalent to each other under certain assumptions. For instance, the tip of a nose curve represents the critical limit for voltage stability which also indicates the power flow insolvability or the singularity of the power flow Jacobian matrix. However, the resulting constraint is only a necessary condition for static voltage stability since the actual system dynamics are not considered. Nevertheless, these analyses based on power flow equations still provide valuable information for system voltage stability and hence being utilized in this paper, as well as many other works, such as \cite{8486639,8728057}.}


\subsubsection{Large signal (transient) voltage stability} \label{sec:2.2.2} 
Due to the complexity and nonlinearity during the transient responses, transient voltage stability is typically assessed through time-domain simulations. In general, it is difficult to mathematically analyze the transient voltage stability and there lacks a standard, reliable and quantitative criterion to determine the transient voltage stability of power systems with high penetration of IBR. Few indices have been proposed to illustrate the transient voltage stability based on post-fault voltage deviation \cite{en14144076,9351059}. Moreover, these indices cannot be calculated analytically and often require time-domain simulations, thus not being able to be conveniently incorporated into system optimization models.

Instead, this work focuses on ensuring enough generation capacity in the system, such that adequate short circuit currents and the post-fault voltages at critical buses after severe faults can be maintained \cite{5271123,NG_GridCode}. Enough SCC ensures that the fault can be cleared by the protection devices within the desired time and proper post-fault voltages prevent further generation tripping, which may lead to other instability and cascade failures. However, it should be noted that enough SCC and adequate post-fault voltage are not a necessary and sufficient condition for transient voltage stability, which is still an open question and is out of the scope of this research. 

Combining the classic SCC superposition approach with the proposed model of IBRs during fault (voltage-dependent current sources) enables the SCC calculation in a general power system with both SGs and IBRs. Based on the superposition principle, the post-fault system operation can be viewed as the superposition of the pre- and `pure'-fault conditions \cite{grainger_stevenson_1994}. The SCC at the fault bus $F$, $I_F^{''}$ can be computed through KCL in the `pure'-fault system where the only sources are the current at the VSC buses and the fault bus \cite{9329077}:
\begin{equation}
\label{-V_F(0)}
    -V_F(0) = \sum_{c\in \mathcal{C}} Z_{F\Phi(c)}(I_{fc}-I_{Lc})+Z_{FF}I_{F}^{''}
\end{equation}
where $V_F(0)$ is the pre-fault voltage at bus $F$; $I_{fc}$ and $I_{Lc}$ is the fault current and load current from IBR $c\in \mathcal{C}$ respectively; $\Phi(c)$ maps the IBR $c\in \mathcal{C}$ to the corresponding bus index. Rearranging \eqref{-V_F(0)} yields the expression of the SCC at bus $F$ as follows:
\begin{equation}
\label{I_sc1}
    I_{F}^{''} = \frac{-V_F(0)-\sum_{c\in \mathcal{C}} Z_{F\Phi(c)}(I_{fc}-I_{Lc})}{Z_{FF}}.
\end{equation}
IBRs are required to provide SCC according to their terminal voltage drop, which can be modeled by voltage-dependent current sources. According to the grid code of UK nationalgridESO \cite{NG_GridCode}, reactive current of full capacity (1.0 - 1.5 $\mathrm{p.u.}$) is required from the GFM IBRs when their terminal voltages drop to zero, to support the post fault voltages and the system protection. Therefore, the fault current from IBR $c\in \mathcal{C}$ can be expressed as follows:
\begin{equation}
\label{I_fc_droop}
    I_{fc} = -\textsf{j} d_c\big({|V_{\Phi(c)}|-|V_{\Phi(c)}(0)|}\big),
\end{equation}
where $d_c \in \mathbb R$ is the reactive current droop gain; $V_{\Phi(c)}$ and $V_{\Phi(c)}(0)$ are the post-fault and pre-fault voltage at bus $\Phi(c)$. Based on the superposition principle, the voltage drop at bus $\Phi(c)$, $\Delta V_{\Phi(c)}$ can be derived:
\begin{equation}
\label{Delta_V}
    \Delta V_{\Phi(c)} = \sum_{c'\in \mathcal{C}} Z_{\Phi(c')\Phi(c)}(I_{fc'}-I_{Lc'})+Z_{F\Phi(c)}I_{F}^{''}.
\end{equation}
Equation \eqref{Delta_V} is essentially an implicit function of $\Delta V_{\Phi(c)}$, due to the dependence of $I_{fc}$ on $\Delta V_{\Phi(c)}$. By combining \eqref{I_fc_droop} and \eqref{Delta_V}, $\forall c\in\mathcal{C}$ and neglecting the pre-fault load current and line resistance, the following expression of $\Delta V_{\mathcal{C}}$ can be obtained:
\begin{equation}
\label{Delta_V_explicit}
    \Delta V_{\mathcal{C}} = A_Z^{-1}
    Z_{F\mathcal{C}}I_{F}^{''},
\end{equation}
where $\Delta V_{\mathcal{C}} \in \mathbb R^{|\mathcal{C}|}$ is the vector of IBR terminal voltage deviations; $ Z_{F\mathcal{C}} \in \mathbb R^{|\mathcal{C}|}$ is the vector collecting $Z_{F\Phi(c)},\,\forall c\in\mathcal{C}$ in the $Z$ matrix; $A_Z\in \mathbb R^{|\mathcal{C}|\times|\mathcal{C}|}$ is defined as:
\begin{equation}
\label{A_Z}
    A_{Z_{ij}}=
    \begin{cases}
    Z_{\Phi(i)\Phi(i)}\textsf{j}d_i+1\;\;&\mathrm{,if}\,i = j \\
    Z_{\Phi(j)\Phi(i)}\textsf{j}d_j\;\;&\mathrm{,if}\,i \neq j
    \end{cases}.
\end{equation}

Combining \eqref{I_sc1}, \eqref{I_fc_droop} and \eqref{Delta_V_explicit} gives the SCC at fault bus $F$:
\begin{equation}
\label{I_sc2}
    I_{F}^{''} = \frac{-V_F(0)+\sum_{c\in\mathcal{C}}{Z_{F\Phi(c)}I_{Lc}}}{Z_{FF}-\textsf{j}Z^{\mathsf T}_{F\mathcal{C}}\mathrm{diag}(d_c)A_Z^{-1}
    Z_{F\mathcal{C}}},
\end{equation}
with $\mathrm{diag}(d_c)\in \mathbb R^{|\mathcal{C}|\times|\mathcal{C}|}$ being the diagonal matrix with $d_c$ being the diagonals. After neglecting the pre-fault load current, the SCC constraint can be expressed as:
\begin{equation}
\label{I_sc3}
    \left|I_{F}^{''} \right|= \frac{\left|V_F(0)\right|}{\left|Z_{FF}-\textsf{j}Z^{\mathsf T}_{F\mathcal{C}}\mathrm{diag}(d_c)A_Z^{-1}
    Z_{F\mathcal{C}}\right|} \ge I_{F_{\mathrm{lim}}}^{''}.
\end{equation}
Note that the above SCC expression reduces to the conventional formula, $V_F(0)/Z_{FF}$, if the SCC from IBR units is neglected, i.e., $d_c=0$.

Moreover, to prevent generation units from disconnecting, the post-fault voltage at their terminals should be maintained above a certain level according to fault ride through requirement of the grid code, for instance, limits of $0.05-0.3\mathrm{p.u.}$ for synchronous units and $0.05-0.15\mathrm{p.u.}$ for IBRs are specified in \cite{Entso_GridCode}. Hence, substituting \eqref{I_sc3} into \eqref{Delta_V_explicit} gives the post-fault voltage constraint at IBR buses in closed form:
\begin{align}
    \label{Delta_V_close}
    \left|\Delta V_{\mathcal{C}} \right| = & \frac{\left|A_Z^{-1}Z_{F\mathcal{C}}V_F(0)\right|}{\left|Z_{FF}-\textsf{j}Z^{\mathsf T}_{F\mathcal{C}}\mathrm{diag}(d_c)A_Z^{-1}
    Z_{F\mathcal{C}}\right|} \le \Delta V_{\mathrm{lim}}.
\end{align}
Remarkably, there are grid codes in other areas with different requirements regarding the fault current contribution from IBR units, such as full capacity reactive current at $0.5\,\mathrm{p.u.}$ and higher terminal voltage drop \cite{ALSHETWI2020119831}. Therefore, the fault current from IBRs as previously defined in \eqref{I_fc_droop} should be modified as:
\begin{equation}
\label{I_fc_droop_1}
    |I_{fc}| = \min\left\{I_c^{\mathrm{max}}, d_c\Delta \left|V_{\Phi(c)} \right| \right\}.
\end{equation}
Moreover, if the pre-fault load current or the line resistance is nonnegligible, a simple explicit expression of the SCC as in \eqref{I_sc2} may not be available in those cases, due to the interdependence between the IBR fault currents and post-fault voltages. However, $I_{F}^{''}$ and $\Delta V_{\mathcal{C}}$ can still be calculated in an iterative manner given the system operating conditions, which has little impact on their formulation as operational constraints, they are discussed in Section~\ref{sec:3}.

\subsection{Frequency Stability} \label{sec:2.3} 
Although frequency stability constraints do not involve system impedance, they closely relate to the status of SGs through the inertia provision, thus influencing system impedance and interacting with other stability constraints. Hence, the frequency stability constraints are also briefly introduced here. We consider the metrics of maximum RoCoF,  steady-state frequency and frequency nadir during the loss of the largest generation, which can be derived based on the Centre-of-Inertia (CoI) model. The first two metrics are simply in linear form, thus not being covered, whereas for the frequency nadir, the following constraint is adopted from \cite{9066910}, where
a novel Wind Turbine (WT) Synthetic Inertia (SI) control scheme is proposed. It eliminates the secondary frequency dip due to WT overproduction and allows the WT dynamics to be analytically integrated into the system frequency dynamics. The frequency nadir constraint is formulated as:
\begin{equation}
\label{nadir_c}
    HR\ge \frac{\Delta P_L^2T_d}{4\Delta f_\mathrm{lim}}-\frac{\Delta P_L T_d }{4} \left(D- \sum_{j\in \mathcal{F}} \gamma_j H_{s_j}^2 \right),
\end{equation}
where $H$ is the sum of conventional inertia and SI of all the wind farms $\sum_{j} H_{s_j}$ with $j\in \mathcal{F}$ being the set of wind farms and $R,\,T_d$ represent the system Primary Frequency Response (PFR) and its delivery time; $\Delta P_L,\, \Delta f_{\mathrm{lim}}$ and $D$ denote system disturbance, limit of frequency deviation and system damping. The last term can be interpreted as: SI provision from each wind farm introduces a negative damping proportional to $H_{s_j}$ with the coefficient being $\gamma_j$, due to their recovery effects \cite{9066910}.

\section{Unified SOC Representation of System Stability Constraints} \label{sec:3}
The stability constraints introduced in the previous section are highly nonlinear, which involve either matrix inverse or eigenvalue operation. Therefore, in order to incorporate these stability criteria into the system scheduling model, a unified reformulation framework that can convert these constraints into optimization-friendly forms is desired. \textcolor{black}{It should also be noted that there exist differences between the stability-constrained optimization and the stability assessment. The former formulates the stability criteria as operational constraints and incorporates them into system-level optimization (the focus of this work), whereas the latter analyzes the system stability based on a known operating condition. Typically, the former establishes the stability constraints based on the analysis from the latter. Having obtained a stability criterion, solving a stability-constrained optimization requires additional effort to transform the stability criterion into a constraint that an optimization model can handle, especially considering the nonlinearity and nonconvexity of the complicated stability analysis. Aiming at determining the optimal system-level decisions that admit the stability constraints, the stability-constrained optimization is much more complicated than assessing the stability of a given operating condition, which is only an evaluation process based on the stability criterion. As a result, the stability representation in the stability-constrained optimization typically cannot achieve the same level of accuracy in terms of system modeling and configurations, which is a sacrifice that has to be made in the stability-constrained optimization. }

Remarkably, the developed system-level stability constraints may involve uncertainties related to system dynamics, stemming from various aspects, such as unknown or inaccurate SG and line impedance, IBR parameters, and system inertia. These uncertainties can be dealt with in the optimization problem by formulating the (distributionally-) robust chance constraints, which are not discussed in this work and will be considered in future studies.

\subsection{General Expression of System Stability Constraints} \label{sec:3.1}
Rewrite the system stability constraints, \eqref{eq:sync_gfl_PQ}, \eqref{ssss_forming}, \eqref{VS} - \eqref{PQ_1}, \eqref{I_sc3}-\eqref{Delta_V_close} and \eqref{nadir_c} derived in Section~\ref{sec:2} in the following form:
\begin{subequations}
\label{gs}
\begin{align}
    &\mathbf{g}_1 (V, Z) = \frac{V^G_{\Phi(c_l)} V_{\Phi(c_l)}}{Z_{\Phi(c_l)\Phi(c_l)}} \ge \cos \phi_Z^G Q_{c_l} + \sin{\phi_Z^G} P_{c_l}  \label{g1}\\
    &\mathbf{g}_2 (V, Z, P_{c_l}) = \lambda_{\mathrm{min}} \left[\mathcal{R}_{\mathcal{C}_m}(Y_{eq})\right] \ge \mathrm{gSCR}_{\mathrm{lim}}  \label{g2}\\
    & \mathbf{g}_3 (V, Z) = \frac{V_F(0)}{Z_{FF}-\sum_{c\in\mathcal{C}}\frac{Z_{F\Phi(c)}^2}{1/d_c + Z_{\Phi(c)\Phi(c)}}} \ge I_{F_{\mathrm{lim}}}^{''} \label{g3} \\
    &\mathbf{g}_4 (V, Z) = \frac{-V_F(0) Z_{F\Phi(c)}}{Z_{FF}\left(1+{d_c}Z_{\Phi(c)\Phi(c)}\right) - \sum_c d_c Z_{F\Phi(c)}^2} \nonumber\\
    & \qquad \quad \;\; \ge -\Delta V_\mathrm{lim} \label{g4} \\
    &\mathbf{g}_5 (\hat{P}_{c_l}, \hat{Q}_{c_l}, \Gamma_{c_l}) = ( \hat{Q}_{c_l}+ \Gamma_{c_l} ) ^ 2 -\hat{P}_{c_l}^2 - \hat{Q}_{c_l}^2 \ge 0 \label{g5}\\
    & \mathbf{g}_6 (H, R, H_s) = HR - \frac{\Delta P_L^2T_d}{4\Delta f_\mathrm{lim}} \nonumber\\
    &\qquad\qquad\quad\;\;-\frac{\Delta P_L T_d }{4} \left(D- \sum_{j\in \mathcal{F}} \gamma_j H_{s_j}^2 \right) \ge 0 \label{g6}
\end{align}
\end{subequations}
where $\hat P_{c_l}$, $\hat Q_{c_l}$ and $\Gamma_{c_l}$ in \eqref{g5} are expressed by the equality constraints as follows:
\begin{subequations}
\label{hs}
\begin{align}
    \mathbf{h}_1 (\hat P_{c_l}, P_{c_l}, Z) &= \hat P_{c_l} - P_{c_l} - \sum_{c_l'\in \mathcal{C},c_l'\neq c_l}\frac{|Z_{\Phi(c_l)\Phi(c_l')}|}{|Z_{\Phi(c_l)\Phi(c_l)}|}P_{c_l'} \nonumber \\
    &=0 \\
    \mathbf{h}_2 (\hat Q_{c_l}, Q_{c_l}, Z) &= \hat Q_{c_l} - Q_{c_l} - \sum_{c_l'\in \mathcal{C},c_l'\neq c_l}\frac{|Z_{\Phi(c_l)\Phi(c_l')}|}{|Z_{\Phi(c_l)\Phi(c_l)}|}Q_{c_l'} \nonumber \\
    &=0\\
    \mathbf{h}_3 (V, Z) &= \Gamma_{c_l} - \frac{{|V^G_{\Phi(c_l)}}|^2}{2|Z_{\Phi(c_l)\Phi(c_l)}|} =0.
\end{align}
\end{subequations}
The inequality constraints defined in \eqref{gs} and equality ones in \eqref{hs} ensure the concerned stability of systems with high IBR penetration. All of them, except the frequency nadir constraint \eqref{g6}, are related to system strength (impedance), which makes them highly nonlinear and involves either matrix inverse or eigenvalue operation with decision-dependence. Hence, it is generally complicated to incorporate them into MILP- or MISOCP-based system scheduling models. In order to achieve this, constraints \eqref{g1} - \eqref{g6} are first expressed in a general form:
\begin{align}
    \label{gi}
    \mathbf{g}_i (\mathsf{X}_i) \ge \mathbf{g}_{i_\mathrm{lim}},\;\;\;i\in \{1,2,3,4,5,6\},
\end{align}
where $\mathsf{X}_i$ is the decision variable in system optimization model and $\mathbf{g}_{i_\mathrm{lim}} \in \mathbb{R}$ can be expressed as:
\begin{subequations}
\begin{align}
    \mathbf{g}_{1_\mathrm{lim}} & = \max\{ \cos \phi_Z^G Q_c + \sin{\phi_Z^G} P_c\} \approx 1 \label{g1_lim} \\
    \mathbf{g}_{2_\mathrm{lim}} & = \mathrm{gSCR}_{\mathrm{lim}}\\
    \mathbf{g}_{3_\mathrm{lim}} & = I_{F_{\mathrm{lim}}}^{''}\\
    \mathbf{g}_{4_\mathrm{lim}} & = -\Delta V_\mathrm{lim}\\
    \mathbf{g}_{5_\mathrm{lim}} & = 0 \\
    \mathbf{g}_{6_\mathrm{lim}} & = 0.
\end{align}
\end{subequations}
The approximation in \eqref{g1_lim} is conservative and it holds because the output of GFL devices during normal operation is dominated by active power and $ \sin \phi_Z^G \gg \cos \phi_Z^G$ in transmission systems. As a result, $\mathbf{g}_{i_\mathrm{lim}},\, \forall i$ are constants regardless of the operating conditions. 

\subsection{Connecting Stability Constraints with Decision Variables} \label{sec:3.2}
Since the elements in the system impedance matrix $Z$ are typically not decision variables in most power system optimization problems, it is necessary to express the impedance matrix with the decision variables in an explicit form, such that the stability constraints can be maintained in the optimization model.

Given the system operating condition, i.e., the status of SGs, $x_{g}$, and the online percentage of GFM $\alpha_{c_{m}}$, the system impedance and admittance matrix in $\mathbf{g}_1$ and $\mathbf{g}_2$ can be calculated by:
\begin{subequations}
\label{Y}
    \begin{align}
    Z &= Y^{-1}\\
    Y &= Y^0 +  Y^g,
    \end{align}
\end{subequations}
where $Y^0$ is the admittance matrix of the transmission lines only; $Y^g$ denotes the additional $Y$ matrix increment due to the reactance of SGs and GFM IBRs. The elements in $Y^g$ can be further expressed as:
\begin{equation}
\label{Y2}
    Y_{ij}^g=
    \begin{cases}
    \frac{1}{X_{g}}x_{g}\;\;&\mathrm{if}\,i = j \land \exists\, g\in \mathcal{G},\, \mathrm{s.t.}\,i=\Psi(g)\\
    \frac{1}{X_{c_{m}}}\alpha_{c_{m}}\;\;&\mathrm{if}\,i = j \land \exists\, c_{m}\in \mathcal{C}_m,\, \mathrm{s.t.}\,i=\Psi(c_{m})\\
    0\;\;& \mathrm{otherwise}.
    \end{cases}
\end{equation}
The above equation manifests the contribution of SGs and GFM units to system voltage strength, since they are modeled as constant voltage sources behind impedances under steady state and small signal disturbances \cite{9282199}. It is further assumed that the voltage magnitudes during normal operation, $V^G_{\Phi(c)}$, $V_{\Phi(c)}$ in \eqref{g1}, \eqref{g2} and $V_F(0)$ in \eqref{g3}, \eqref{g4} equal to $1\,\mathrm{p.u.}$. Hence, $\mathbf{g}_1$ becomes a function of $x_g$ and $\alpha_{c_{m}}$, i.e., $\mathsf{X}_1=[x_1,...x_g,...x_{|\mathcal{G}|}, \alpha_{c_1},...,\alpha_{c_{m}}...,\alpha_{|\mathcal{C}_{m}|}]$. Additionally, $\mathbf{g}_2$ also depends on the output power of GFL IBRs as defined in \eqref{Yeq}. As a result, $\mathsf{X}_2$ can be written as $[x_1,...x_g,...x_{|\mathcal{G}|}, \alpha_{c_1},...,\alpha_{c_{m}}...,\alpha_{|\mathcal{C}_{m}|}, P_{c_1},...,P_{c_{l}}...,P_{|\mathcal{C}_{l}|}]$.

On the other hand, during transient processes, both GFM and GFL IBRs are controlled to inject reactive current depending on their terminal voltages, which are modeled as voltage-dependent current sources through the droop control \eqref{I_fc_droop}. As a result, the only voltage sources in the system are the SGs. Therefore, the matrix $Y^g$ used to calculate $\mathbf{g}_3$ and $\mathbf{g}_4$ includes the impedances from SGs only, and the contribution on system SCC and post-fault voltage from IBRs is modeled through the droop control. To further account for the variation of IBR online capacity due to the renewable resource variation, the IBR droop control gain, $d_c$, is replaced by $\alpha_c d_c$ with the coefficient $\alpha_{c},\,c\in \mathcal{C}$ representing the online percentage of IBR unit $c$. Therefore, $\mathsf{X}_3$ and $\mathsf{X}_4$ can be expressed as, $\mathsf{X}_i=[x_1,...x_g,...x_{|\mathcal{G}|}, \alpha_{1},...,\alpha_{c}...,\alpha_{|\mathcal{C}|}],\, {i\in\{3,4\}}$. As for $\mathbf{g}_5$ and $\mathbf{g}_6$, they are explicit functions of the decision variables, i.e.,  $\mathsf{X}_5 = [\hat{P}_{c_l}, \hat{Q}_{c_l}, \Gamma_{c_l}]$ and $\mathsf{X}_6 = [H,R,H_s]$.

It should be noted that $x_{g},\,\forall g \in \mathcal{G}$ can be viewed as binary decision variables, whereas for the IBRs, their operating conditions are determined by the current available wind/solar resources rather than the system operator. Therefore, the scenario-dependent parameter $\alpha_{c}\in [0,1],\,\forall c \in \mathcal{C}$ is introduced to represent the percentage of IBRs' online capacity. The approach proposed in \cite{6672214} and \cite{7370811} is used in this paper where the online IBR capacity is estimated given the current available power based on historical data. Although there are also some direct approaches to monitor or estimate the online capacity of the IBRs and could potentially provide more accurate results, it is out of the scope of this paper and is not discussed in more detail.

\subsection{SOC Representation of Stability Constraint Boundary}\label{sec:3.3}
\color{black}
Conventionally, optimization problems in power systems varying from economic dispatch, unit commitment, optimal power flow to market-clearing, and system planning are commonly formulated into a (Mixed-Integer) Linear Programming [(MI)LP]. However, due to the increasing IBR penetration and uncertainty level in power systems, these MILP-based optimizations present limitations in different aspects. First, the classical methods within the MILP framework, such as scenario-based stochastic programs and robust optimization techniques may suffer from computational intractability and solution conservatism, respectively. Second, an accurate network modeling may not be achievable using DC power flow or linearized AC power flow. Third, the security and stability issues brought by high IBR penetrations may not be representable with linear structures.

As a result, the LP-based optimization problems in power systems are augmented with conic constraints to mitigate these issues to different extents. In the area of uncertainty modeling and management, (distributionally robust) chance-constrained optimizations, which admit nonlinear yet convex, computationally tractable and analytically expressable uncertainty models have been recently studied. As for more accurate modeling of the network constraints, the SOCP-based AC power flow relaxation has also been proposed. Regarding the security and stability-constrained optimization, which is also the focus of this work, using SOC constraints to represent and approximate the highly nonlinear stability constraints has also shown promising results.

After reformulating the stability constraints into SOC form, the resulting SOCP-based optimization problems can be further efficiently solved in polynomial time using interior-point methods by several off-the-shelf commercial solvers such as MOSEK, Gurobi, and CPLEX \cite{ratha2023moving}. In the following, a unified framework that can effectively reformulate the stability constraints, in a general way, to fit any typical power system optimization model is presented.
\color{black}

It is understandable that due to the complex expression of $\mathbf{g}_i$ and especially their dependence on the discrete variables, $x_g$ through the matrix inverse and eigenvalue operations, it is generally difficult to deal with those constraints in an optimization problem. The target of the stability constraint reformulation is to describe the boundary of the stability feasible region through an SOC relationship. Although the SOC form is chosen because it is the most general constraint whose mixed-integer programming can be efficiently solved by commercial solvers, the proposed method can also be applied to obtain constraints in linear or even semidefinite form. 

The estimated expressions of the nonlinear functions $\mathbf{g}_i$ can be defined as follows:
\begin{align}
    \label{g_SOC}
    \Tilde{\mathbf{g}}_i  (\mathsf{X}_i) & = \mathsf{c}_i \mathsf{X}_i+ \mathsf{d}_i - \left\lVert \mathsf{A}_i \mathsf{X}_i+ \mathsf{b}_i \right\lVert,
\end{align}
where $\Tilde{\mathbf{g}}_i$ is the estimated function of $\mathbf{g}_i$ in SOC form and the matrices $\mathsf{A}_i\in\mathbb{R}^{j\times \dim( \mathsf{X}_i)}$, $\mathsf{b}_i\in\mathbb{R}^{j}$, $\mathsf{c}_i\in\mathbb{R}^{1\times\dim( \mathsf{X}_i)}$ and $\mathsf{d}_i\in\mathbb{R}$ are parameters to describe the shape and position of the SOC. The value of $j$ is a degree of freedom that should be chosen for the best reformulation. A larger $j$ may yield a more accurate description of the original nonlinear relationship, but result in more parameters and computational burden. By substituting \eqref{g_SOC} into \eqref{gi}, the system stability constraints can also be converted to SOC form straightforwardly:
\begin{align}
    \label{SOC_glim}
    \left\lVert \mathsf{A}_i \mathsf{X}_i+ \mathsf{b}_i \right\lVert \le \mathsf{c}_i \mathsf{X}_i+ \left( \mathsf{d}_i - \mathbf{g}_{i_\mathrm{lim}} \right).
\end{align}

In order to find the SOC parameters $\mathcal{K} =\{\mathsf{A}_i, \mathsf{b}_i, \mathsf{c}_i, \mathsf{d}_i\},\,\forall i$ that can best describe the boundary of the system stability feasible region, the following boundary-aware optimization is proposed:
\begin{subequations}
\label{DM3}
\begin{align}
    \label{obj3}
    \min_{\mathcal{K}}\quad & \sum_{\omega^i \in \Omega^i_2} \left(\mathbf{g}_{i}^{(\omega^i)} - \Tilde{\mathbf{g}}_{i}^{(\omega^i)} \right)^2\\
    \label{coef_ctr2}
    \mathrm{s.t.}\quad & \Tilde{\mathbf{g}}_{i}^{(\omega^i)}< \mathbf{g}_{i_{\mathrm{lim}}},\,\,\forall \omega^i \in \Omega^i_1\\
    \label{coef_ctr3}
    &\Tilde{\mathbf{g}}_{i}^{(\omega^i)} \ge {\mathbf{g}}_{i_{\mathrm{lim}}},\,\,\forall \omega^i \in \Omega^i_3,
\end{align}
\end{subequations}
with $\omega^i = \{\mathsf{X}_i^{(\omega^i)},\, \mathbf{g}_{i}^{(\omega^i)}\}\in \Omega^i$ denoting the entire data set corresponding to stability constraint $i$. It is generated by evaluating $\mathbf{g}_{i}$ in representative system conditions $\mathsf{X}_i$. For the SGs, all the possible generator combinations are considered, whereas for the continuous parameters $\alpha_{c}\in [0,\,1]$, there are infinite possible conditions. To obtain the data set with a finite size, the interval is evenly divided into $n_c$ regions, $\forall c\in \mathcal{C}$, each of which is represented by its mean value. Hence, the total size of $\Omega^i$ is $2^{|\mathcal{G}|}\cdot n_c^{|\mathcal{C}|}$. Moreover, this number can also be significantly reduced by considering the actual operating conditions in the system. 
The sets $\Omega^i_1 ,\, \Omega^i_2$ and $\Omega^i_3 $ are the subsets of $\Omega^i$, whose relationship is defined as below:
\begin{figure}[!b]
    \centering
    \vspace{-0.3cm}
	\scalebox{0.2}{\includegraphics[trim=0 0 0 0,clip]{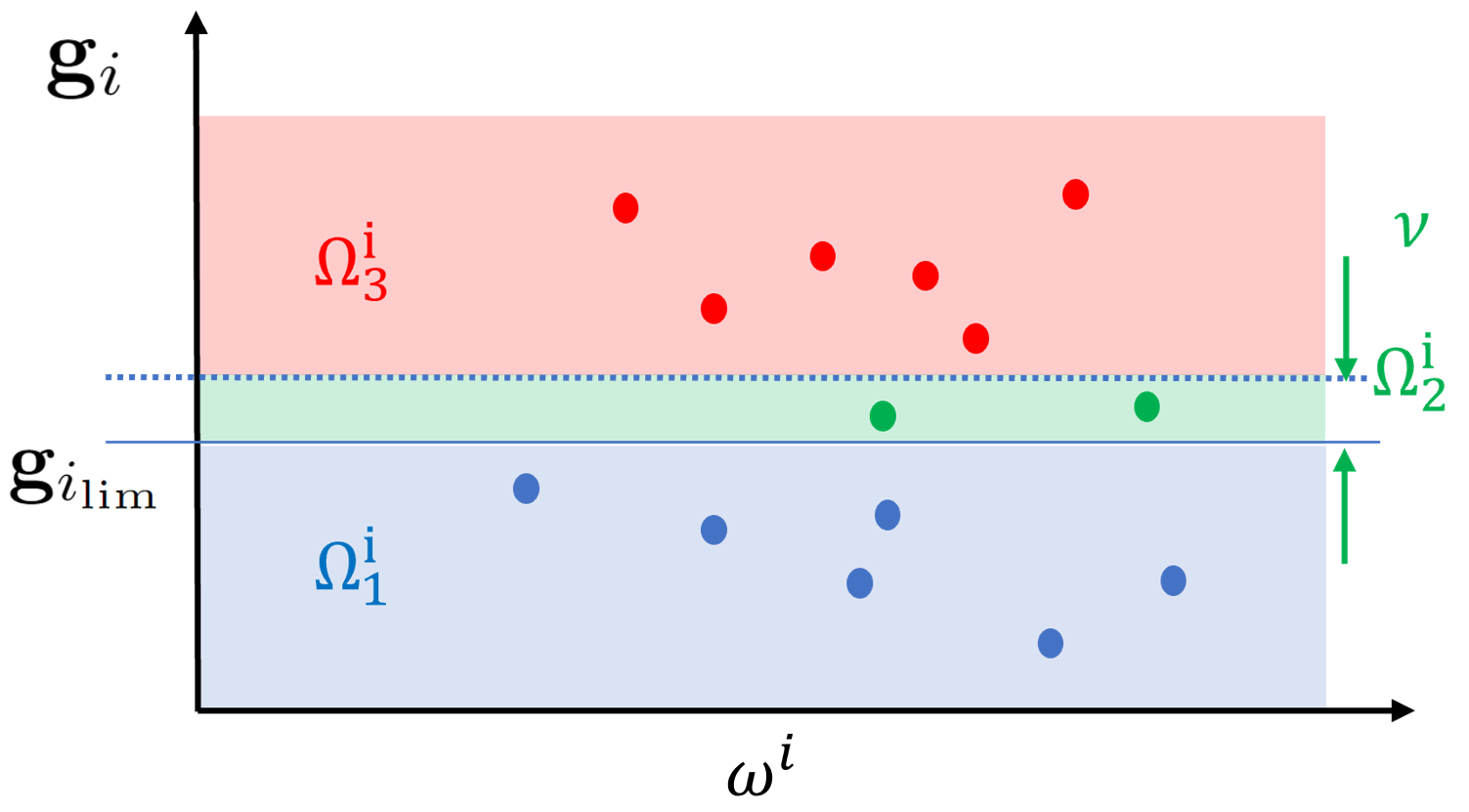}}
    \caption{\label{fig:Omega}Relationship of different data set.}
\end{figure}
\begin{subequations}\label{Omega}
\begin{align}
    \Omega^i &= \Omega^i_1 \cup\Omega^i_2\cup\Omega^i_3 \\
    \label{Omega1}
    \Omega^i_1 & = \left\{\omega^i\in \Omega^i \mid \mathbf{g}_{i}^{(\omega^i)}<\mathbf{g}_{i_{\mathrm{lim}}} \right\}\\
    \label{Omega2}
    \Omega^i_2 & = \left\{\omega^i\in \Omega^i \mid \mathbf{g}_{i_{\mathrm{lim}}} \le \mathbf{g}_{i}^{(\omega^i)}<\mathbf{g}_{i_{\mathrm{lim}}} + \nu \right\}\\
    \label{Omega3}
    \Omega^i_3 & = \left\{\omega^i\in \Omega^i \mid \mathbf{g}_{i_{\mathrm{lim}}} + \nu\le \mathbf{g}_{i}^{(\omega^i)} \right\},
\end{align}
\end{subequations}
with $\nu$ being a constant parameter. An example of the relationship in \eqref{Omega} is demonstrated in Fig.~\ref{fig:Omega}. Given \eqref{coef_ctr2} and \eqref{Omega1}, all the data points whose real stability indices, $\mathbf{g}_{i}^{(\omega^i)}$ are smaller than the limits, can be identified correctly by the estimated function, $\Tilde{\mathbf{g}}_{i}^{(\omega^i)}$. Ideally, it is also desired to correctly identify all the above-limit data points, which would make the problem become a classification model. However, this may cause infeasibility due to the restricted SOC structure defined in \eqref{g_SOC}. Therefore, a parameter $\nu\in \mathbb{R}^+$ is introduced to define $\Omega^i_2$ and $\Omega^i_3$ as in \eqref{Omega2} and \eqref{Omega3}. In this way, all the data points in $\Omega^i_3$ (red area) will be classified correctly and misclassification can only occur in $\Omega^i_2$ (green area), thus being conservative. 

Furthermore, $\nu$ should be chosen as small as possible while ensuring the feasibility of \eqref{DM3}. Note that only the errors of data points in $\Omega^i_2$ are penalized in the objective function \eqref{obj3}, since for the data points in $\Omega^i_1$ and $\Omega^i_3$, as long as they are classified on the correct side of the limits, the errors are of no concern. As a result, the proposed method ensures accurate approximation within a narrow region around the limits ($\Omega^i_2$) whereas the regression errors of the data points in the other two regions are insignificant. Note that although \eqref{coef_ctr3} is a constraint in SOC form, \eqref{coef_ctr2} is nonconvex and needs to be linearized. This may increase the computational burden of solving \eqref{DM3}, however, since it is an offline process, the computational time is not an issue. 

As the original relationship in $\mathbf{g}_5$ and $\mathbf{g}_6$ resembles to that of an SOC, the parameters $\mathcal{K}_i =\{\mathsf{A}_i, \mathsf{b}_i, \mathsf{c}_i, \mathsf{d}_i\},\, i \in \{5,6\}$ can be directly identified by transforming $\mathbf{g}_i \ge \mathbf{g}_{i_\mathrm{lim}}$ into SOC form. For $\mathbf{g}_5 \ge \mathbf{g}_{5_\mathrm{lim}}$, it can be converted to standard SOC form as follows:
\begin{align}
    \left\lVert \begin{bmatrix}\hat P_c\\ \hat Q_c\end{bmatrix} \right\rVert_2  \le \hat{Q}_c +\Gamma_c.
\end{align}
The parameters $\{\mathsf{A}_5, \mathsf{b}_5, \mathsf{c}_5, \mathsf{d}_5\}$ can therefore be determined accordingly. For $\mathbf{g}_6 \ge \mathbf{g}_{6_\mathrm{lim}}$, the method proposed in \cite{9475967} is used to convert the frequency nadir constraint $\mathbf{g}_6$ to SOC form. The original constraint $\mathbf{g}_6 \ge \mathbf{g}_{6_\mathrm{lim}}$ can be rewritten as follows:
\begin{equation}
\label{nadir_soc}
    HR\ge \underbrace{\frac{\Delta P_L^2T_d}{4\Delta f_\mathrm{lim}}-\frac{\Delta P_L T_d D_0 }{4}}_{x_1^2} + \frac{\Delta P_L T_d \sum_{j\in \mathcal{F}} \gamma_j H_{s_j}^2}{4} .
\end{equation}
Being a constant, the ancillary variable $x_1$ is defined for the sole purpose of SOC reformulation. Additionally, since in real power systems, the relationship $\Delta P_L/\Delta f_\mathrm{lim}>D_0$ always holds, $x_1$ is real and \eqref{nadir_soc} is well defined, which enables to express the nadir constraint in SOC form as below:
\begin{align}
\label{nadir_SOC_standard}
    \left\lVert \begin{bmatrix}2x_1\\ \sqrt{\Delta P_L T_d} \mathbf{H_s}\\ H-R \end{bmatrix} \right\rVert_2  \le H + R
\end{align}
where $\mathbf{H_s}=\begin{bmatrix} \sqrt{\gamma_1} H_{s_1}& \sqrt{\gamma_2} H_{s_2}&...& \sqrt{\gamma_{|\mathcal{F}|}} H_{s_{|\mathcal{F}|}} \end{bmatrix}^\mathsf{T}$. Based on \eqref{nadir_SOC_standard}, the SOC parameters $\{\mathsf{A}_6, \mathsf{b}_6, \mathsf{c}_6, \mathsf{d}_6\}$ can be identified.

\subsection{Linear Reformulation of Equality Constraints}
It can be observed that the only nonlinear terms in \eqref{hs} are those related to the impedance matrix, ${|Z_{\Phi(c)\Phi(c')}|}/{|Z_{\Phi(c)\Phi(c)}|}$ and ${1}/{|Z_{\Phi(c)\Phi(c)}|}$. However, different from the inequality constraints \eqref{gi}, where accurate approximation is only required in the area around the boundaries, the linearization of the equality constraints \eqref{hs} is desired to be accurate for all the operating conditions. Hence, for the equality constraints, the method proposed in Section~\ref{sec:3.3} does not apply, since the boundary to be identified does not exist. Instead, a standard least square regression is utilized to find the best linear expression that describes the relationship between ${|Z_{\Phi(c)\Phi(c')}|}/{|Z_{\Phi(c)\Phi(c)}|}$, ${1}/{|Z_{\Phi(c)\Phi(c)}|}$ and the decision variables $x_g$, $\alpha_{c_m}$ \cite{9786660}. 

Taking $\mathbf{h}_1 = 0$ for instance, define its linearized expression as follows:
\begin{align}
\label{P_linear}
    \hat P_{c_l} - P_{{c_l}}& -  \sum_{{c_l'}\in \mathcal{C},{c_l'}\neq {c_l}}  \left(\sum_{g\in\mathcal{G}} \mathsf{k}_{g}^{{c_l'}} x_{g} + \sum_{c_m\in\mathcal{C}_m} \mathsf{k}_{c_m}^{{c_l'}} \alpha_{c_m} \right. \nonumber \\
    & + \left.\sum_{m\in\mathcal{M}} \mathsf{k}^{{c_l'}}_{m} \eta_m \right) P_{{c_l'}}=0,
\end{align}
where $\mathsf{K} =\{\mathsf{k}_{g}^{{c_l'}},\,\mathsf{k}_{c_m}^{{c_l'}},\,\mathsf{k}_{m}^{{c_l'}}\},\,\forall g,c_m,m$ are the sets of linear coefficients corresponding to ${|Z_{\Phi(c)\Phi(c')}|}/{|Z_{\Phi(c)\Phi(c)}|}$ determined by solving the least square regression within the entire data set. To further account for the nonlinearity in the original relationship, the term $\eta_m$ is added to describe the interactions between every two units in SGs and GFM IBRs, i.e., $m\in\mathcal{M} =\{m_1,\,m_2 \mid m_1,\,m_2\in \mathcal{G} \cup \mathcal{C}_m\}$, which is defined as follows:
\begin{align}
    \label{eta}
    \eta_m  =
    \begin{cases}
        x_{m_1}x_{m_2},\quad \mathrm{if} \,\,m_1,m_2 \in \mathcal{G} \\
        x_{m_1}\alpha_{m_2},\quad \mathrm{if} \,\,m_1\in \mathcal{G},\,m_2 \in \mathcal{C}_m \\
        \alpha_{m_1}\alpha_{m_2},\quad \mathrm{if} \,\,m_1,m_2 \in \mathcal{C}_m
    \end{cases},\,\,\forall m \in \mathcal{M}.
\end{align}
Note that since the 2nd order terms lead to accurate enough results, the higher-order terms are neglected in order to achieve a balance between accuracy and computational effort. 

\section{Conclusion} \label{sec:4}
This paper proposes a stability-constrained optimization framework in high IBR-penetrated power systems to ensure system stability and security. System stability constraints covering from IBR synchronization and voltage stability to frequency stability are developed analytically. For the IBR synchronization stability, system operational constraints of both equilibrium point existence and small-signal stability are derived. Voltage stability involves small-signal and transient stability constraints, the latter of which is characterized by the sufficiency of generating capacity to provide fault current and maintain post-fault voltages.

Depending on system impedance, generator status and operating point, the highly nonlinear stability constraints are further reformulated into SOC form through a unified boundary-aware data-driven approach to achieve accuracy and conservativeness. The resulting constraints can be incorporated into any general power system optimization model. Part II of this paper validates the developed constraints and investigates their potential application in system scheduling and stability market design. 

\bibliographystyle{IEEEtran}
\bibliography{bibliography}
\end{document}